# Groundwater vulnerability assessment in semi-arid regions using GIS-based DRASTIC models and FUZZY AHP: South Chott Hodna

Lakhdar Seraiche [a], Mostafa Dougha [a,*], Messaoud Ghodbane [a], Tahar Selmane [a,b], Ahmed Ferhati [a], Djamal Eddine Djemiat [a]

[a] *VEHDD Laboratory, Faculty of Technology, University of M'sila, Univ. Pole, Road of BBA, M'sila 28000, Algeria*
[b] *Department of Hydraulics and Civil Engineering, University of Ghardaia, Algeria*

## HIGHLIGHTS

- Model Enhancement through Land Use Integration: The traditional DRASTIC model was improved (DRASTIC_LU) by incorporating land use data, offering a more realistic assessment of anthropogenic pressures on groundwater in Algeria's semi-arid South Chott Hodna region.
- Use of Advanced Multi-Criteria Tools: Analytical Hierarchy Process (AHP) and Fuzzy AHP were applied to optimize parameter weighting and address uncertainty, enhancing decision support in vulnerability mapping.
- Strong Predictive Performance: The enhanced models demonstrated high accuracy in linking nitrate contamination to vulnerability.

## GRAPHICAL ABSTRACT

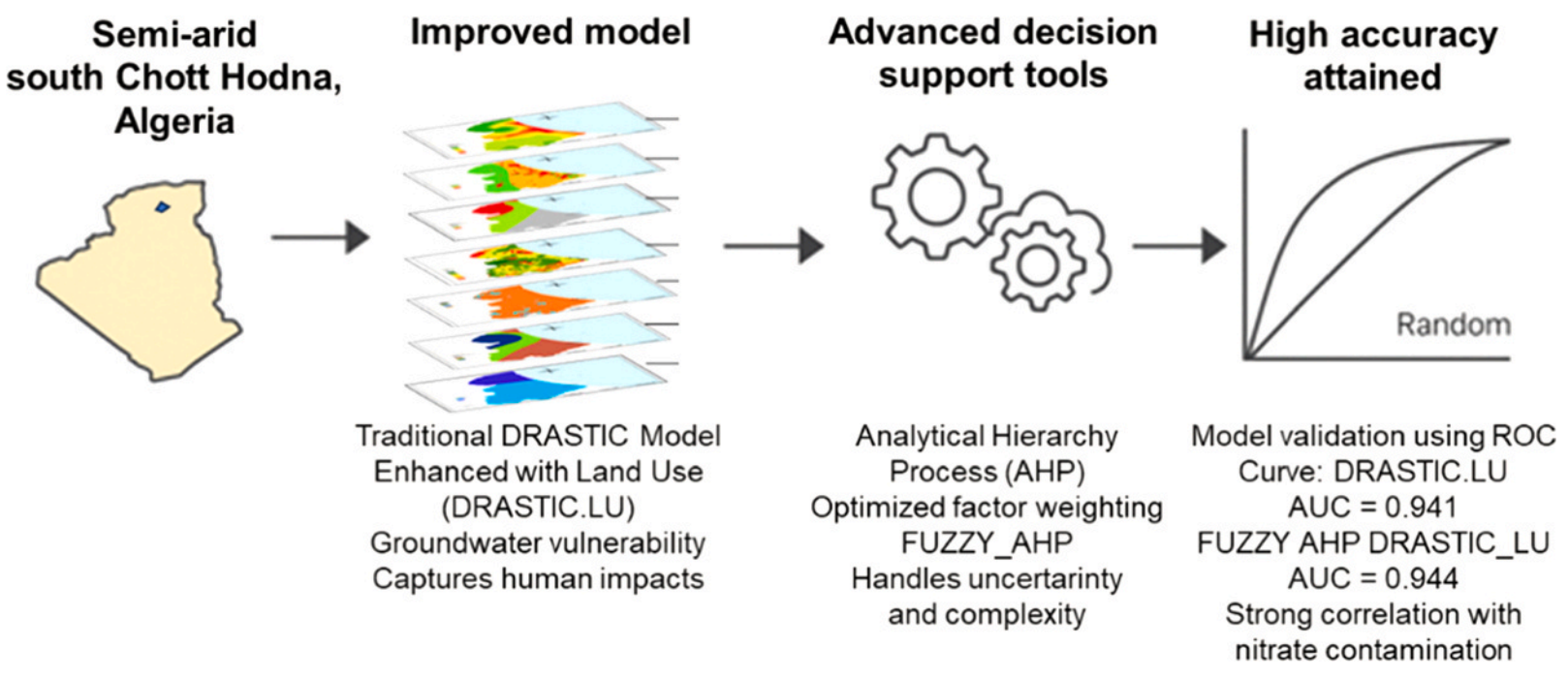

## ARTICLE INFO

## ABSTRACT

Groundwater vulnerability is a major concern in arid regions worldwide, where population growth and intensive agriculture increase the risks of depletion and contamination. This study proposes a hybrid groundwater vulnerability assessment framework that improves the conventional DRASTIC model by integrating land-use data and applying advanced weighting techniques, namely the Analytical Hierarchy Process (AHP) and its fuzzy logic variant (Fuzzy AHP). This method makes expert-based weighting less subjective, better captures anthropogenic effects, and facilitates adaptation to challenging situations and limited data. Four vulnerability maps were produced using Geographic Information Systems (GIS): DRASTIC, DRASTIC_LU, AHP DRASTIC_LU, and Fuzzy AHP DRASTIC_LU. We used nitrate levels from 70 wells to verify our work. We found that agricultural areas, especially those above the alluvial aquifer, were the most vulnerable. The ROC curve analysis showed that the model improved over time, with the area under the curve (AUC) values of 0.812 for DRASTIC, 0.864 for DRASTIC_LU, 0.875 for AHP DRASTIC_LU, and 0.951 for fuzzy AHP DRASTIC_LU. These results show that fuzzy AHP DRASTIC_LU makes groundwater risk assessments much more. The GIS-based hybrid models offer a scalable and transferable method for mapping vulnerability, but they also provide local and regional water resource managers with useful information.

* Corresponding author.
*E-mail address:* mostafa.dougha@univ-msila.dz (M. Dougha).

## 1. Introduction

Groundwater is the main source of water for farming and homes in the semi-arid Chott Hodna Sud region of Algeria, where surface water resources are limited. But rising pressures from climate change, urbanization, and intensive farming are speeding up the depletion of groundwater and raising the risk of contamination. Nitrate pollution has become a significant problem in southern Chott Hodna, particularly in the Maadher Plain and along Wadi Maitre and Bou Saada, over the last 20 years. This pattern is similar to what has been seen in agricultural areas around the world, where the overuse of chemical fertilizers, animal waste, and untreated wastewater are the main causes of groundwater degradation [1,2]. Recent field surveys in our study area (August–September 2023) showed that nitrate levels in groundwater often went above the World Health Organization's (WHO) drinking water limit of 50 mg/L [3]. Some wells even had levels as high as 300 mg/L. These results indicate a need for robust mapping of groundwater vulnerability that can accurately identify high-risk areas and assist us in developing strategies to mitigate the risk.

To deal with these problems, tools for assessing groundwater vulnerability need to include both the physical features of aquifers and the changing human-made pressures that affect the risk of contamination. The DRASTIC index is still one of the most popular models because it is easy to use, works with GIS, and includes important hydrogeological parameters like depth to water, net recharge, aquifer and soil media, slope, vadose zone impact, and hydraulic conductivity [4]. The model's use of fixed weights in different geographic areas, on the other hand, makes it less able to respond to changes in the environment and human activity. In response, many researchers have suggested hybrid models that improve DRASTIC by using Multi-Criteria Decision Analysis (MCDA) tools like the Analytical Hierarchy Process (AHP), which lets you change the weight of factors over time. Bera et al. (2022) and Saravanan et al. (2023) used AHP-enhanced DRASTIC in Indian semi-arid areas and found that it was better at predicting vulnerability than the original model [5,6]. Chakraborty et al. (2022) and Bera et al. (2021) also stressed the importance of taking into account land use and agricultural contamination in similar situations [7,8].

Other recent studies have broadened the field of groundwater assessment by using GIS models with classification algorithms or fuzzy logic. For example, Derdour et al. (2023) showed that machine learning-based vulnerability modeling works well in Algeria's dry areas [9]. Kaliyappan et al. (2024) came up with a fuzzy-weighted method to make groundwater suitability analysis better when there isn't much data available [10]. Alhamd and Ibrahim (2024) also showed that using spatial modeling to look at how salinity changes over time in relation to farming practices can help us understand how to keep water in dry areas [11]. Alsaeed et al. (2025) looked at surface water treatment processes, but their use of response surface modeling supports the growing agreement that combining spatial tools with analytical models makes it easier to make decisions in water systems [12]. Even with these improvements, there aren't many studies that combine land use data, fuzzy logic, and AHP into a single vulnerability mapping framework, especially in North African semi-arid areas. Even fewer studies check their models against field-measured contamination indicators.

In this context, the current study presents a new hybrid framework that improves the traditional DRASTIC model by: (i) adding land use as a contamination risk layer (DRASTIC_LU), (ii) using AHP and fuzzy AHP to optimize parameter weights, and (iii) checking results against actual nitrate levels. This integrated approach fixes major problems with current models, like fixed weights and no field verification. It also gives us a strong, flexible tool for assessing groundwater risk and planning for sustainable resource use in dry and semi-dry areas.

## 2. Study area

The study area is in the southern part of the Chott Hodna basin, southeast of Algiers, between latitudes 35°7′ and 35°28′N and longitudes 4°6′ and 4°48′E. It is about 450 m above sea level on average (Fig. 1). It is about 112,730 ha big and is bordered by Chott Hodna to the north, Mount Meharga to the south, Kerdada Peak and Mount Aouidja to the southwest, Mount Baieun and Mount Arar to the west, and Oued M'cif to the east. The climate in this area is changing from dry to the semi-arid. The winters are usually mild and wet, while the summers are hot and dry. There isn't much rain all year, which makes water even scarcer and makes the area more vulnerable to hydrological stress. Loose alluvial deposits and sand dunes also contribute to wind erosion and surface instability. The area traverses four major wadis: Bousaada, Romana, Maîter, and M'cif. While Bousaada and Romana terminate in closed depressions, Maîter and M′cif discharge directly into Chott Hodna during flood events. These episodic floods play a critical role in aquifer recharge, especially in zones where floodwater infiltration through permeable wadi beds is prevalent.

There is a thick aquifer system below the surface of the region that goes back to the Mio-Pliocene to Quaternary period and reaches depths of up to 250 m [13,14]. There are layers of clay, sandy clay, sandstone, and clayey conglomerates in this system. The main aquifers are made up of sandy and conglomerate formations from the Mio-Pliocene. These are connected to deeper Cretaceous units through fractured interfaces (Fig. 2), which make it easier for groundwater to flow up and down and sideways [15]. Recharge happens in three main ways: direct rainfall infiltration, floodwater infiltration through wadis, and return flow from irrigation practices. All of these can be ways for pollutants to move from the surface to the saturated zone [8,16].

Groundwater levels have been steadily dropping since the 1970s, which means that extraction pressure is rising and recharge is not lasting long enough [17]. Also, a lot of farming has been done on sandy clays that are brown to reddish-brown. These soils are good for growing crops, but they have also led to unregulated drilling of wells that are 90–220 m deep, often without proper sealing. These actions greatly raise the risk of direct contamination from surface sources, such as fertilizers, animal waste, and wastewater.

Because of its hydrogeological vulnerability and the pressures people put on it, the study area is a wonderful place to use an enhanced vulnerability mapping approach. The DRASTIC_LU model is a suitable fit here because it takes into account both the natural properties of the aquifer and how the land is used, both of which have a big impact on the risk of contamination. This model, when used with MCDA methods like AHP and fuzzy AHP, makes it possible to assess groundwater vulnerability in a more realistic and flexible way in semi-arid areas where there isn't much data.

## 3. Materials and methods

### 3.1. Data collection

This study used both primary and secondary data to help with spatial modeling and vulnerability assessment. We got meteorological and hydrogeographic data from the Algerian Regional Directorate of Water Resources. This included records of rainfall, information about the river network, and piezometric measurements. Between August and September 2023, a groundwater sampling campaign took place. It included 70 wells spread out over the study area. We put the samples in sterilized polyethylene bottles to keep their chemical integrity and keep them from getting contaminated. The accredited Algerian water laboratory used spectrometric methods to quickly and accurately verify the nitrate ($NO_3^-$) levels right after the samples were taken. Adding field-verified nitrate levels made it possible to effectively examine the vulnerability maps that came out of the study. We used well logs, borehole lithology surveys, and piezometric survey data, as well as reliable data, to figure out the characteristics of the aquifer and the levels of groundwater. We used remote sensing images to make land cover and land use (LU) maps, which we then checked against field observations and administrative databases.

### *3.2. Preparation of thematic layers*

To evaluate groundwater vulnerability, we used a GIS platform to integrate and standardize nine thematic layers representing hydro-environmental factors in the DRASTIC and DRASTIC_LU models. These include depth to water (D), net recharge (R), aquifer media (A), soil media (S), topography (T), impact of the vadose zone (I), hydraulic conductivity (C), and land use (LU). We used ArcGIS 10.8 to turn all the parameters into raster format. Depending on the density of the data and its spatial correlation, we used Inverse Distance Weighting (IDW) or Ordinary Kriging to fill in the gaps. Using the Jenks natural breaks method, each layer was put into one of five vulnerability classes. This made sure that the results were consistent and statistically representative.

The depth to the water table (D), which is the vertical distance pollutants have to travel to get to the aquifer, was calculated using data from 19 wells with water levels between 5.70 and 66.50 m. The IDW interpolation method was used, and the parameter got a score between 2 and 10, with 5 being the highest.

Taking into account rainfall, lateral infiltration, wadi inflow, and irrigation return flows, the net recharge (R) was thought to be between 42 and 390 mm/year. We used kriging to fill in the gaps in this layer and gave it a weight of 4.

The aquifer media (A) layer was made from piezometric and borehole logs that showed the rock type at each sampled well, such as limestone, sandstone, clay, or gravel. The samples were taken from depths of 25–95 m, which included both shallow and deeper confined layers of the Maadher Plain aquifer system. Using kriging and a weight of 3, this important parameter for figuring out how contaminants spread was interpolated.

Soil media (S), which controls how contaminants move through the surface by allowing them to pass through and stick to things, was taken from geological profiles and put into five main groups (e.g., sandy clay loam, gravel-rich clay). Finer materials such as silt and clay reduce pollutant mobility through adsorption, whereas coarser soils enhance infiltration. This parameter was weighted at 2.

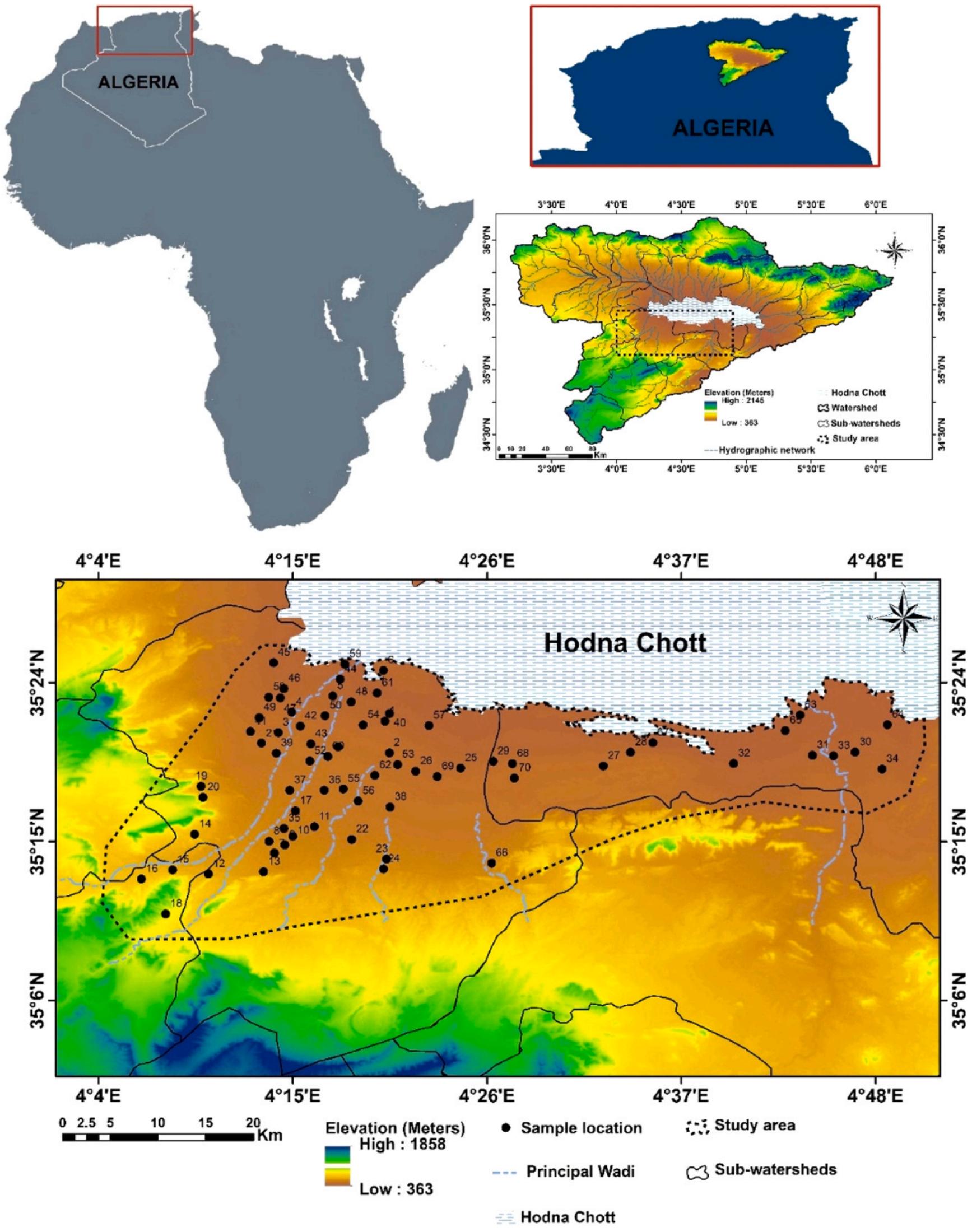

**Fig. 1.** Geographic location and sampling of the study area.

Topography (T), influencing surface runoff and infiltration potential, was derived from an SRTM-based Digital Elevation Model (resampled from 30 ×30 m to 10 ×10 m). The slope ranged from 0 % to 40.5 %, classified into five categories, and assigned a low vulnerability weight of 1 in accordance with the original DRASTIC model.

The vadose zone (I) was mapped using lithological data between the ground surface and water table. The materials included sandy loam and dense clays, which affected how pollutants flowed vertically and how quickly they broke down. We used kriging to fill in this parameter and gave it a high vulnerability weight of 5. It controls how well contaminants can move through unsaturated layers.

We used transmissivity and aquifer thickness data from boreholes to figure out the hydraulic conductivity (C), which controls how contaminants spread horizontally once they get into the aquifer. We used kriging to fill in the gaps in the conductivity values and gave them a weight of 3. Zones with high permeability indicated a higher likelihood of contamination.

Land use (LU) was added as an extra layer to the DRASTIC_LU model to take into account human activity. Using remote sensing and GIS, we divided the area into five main types of land use: urban areas, old and new irrigated agriculture, areas affected by septic tanks and sewage, and empty land. The scores went from 1 (least polluting) to 10 (most polluting), with a weight of 5. This shows how much land use affects nitrate infiltration in this semi-arid basin.

Following standard DRASTIC and regional adaptations [8,18–21], all thematic factors were given a score from 1 (low vulnerability) to 10 (high vulnerability). Based on WHO drinking water guidelines from 2017, the nitrate threshold for model validation was set at 50 mg/L. This threshold was used to see how well vulnerability maps could predict what would happen in the real world. A summary of all weights, scoring ranges, and layer-specific values are provided in Table 1.

### 3.3. Methodology

To assess groundwater vulnerability and contamination risk from human activities in the southern Chott Hodna region, we applied and improved the conventional DRASTIC model by integrating land use data (DRASTIC_LU) and optimizing factor weightings using decision support tools. The analysis followed a structured sequence: (1) seven hydrogeological parameters (water depth, net recharge, aquifer environments, soil environments, topography, vadose zone impact, and hydraulic conductivity) were mapped as GIS layers; (2) an additional land use layer was integrated to account for anthropogenic influences; (3) factor assessments and weightings were determined using the Analytical Hierarchy Process (AHP) and its fuzzy logic variant (Fuzzy AHP) to reduce subjectivity in weighting assignment; (4) vulnerability indices were calculated using a weighted overlay in a GIS environment; and (5) model results were validated using observed nitrate concentrations by Receiver Operating Characteristic (ROC) curve analysis. This integrative approach enabled the production of high-resolution vulnerability maps and improved the reliability of contamination risk assessment in semi-arid regions.

Many researchers have proposed hybrid models that enhance DRASTIC using Multi-Criteria Decision Analysis (MCDA) tools like the Analytical Hierarchy Process (AHP), which allow for more flexible and expert-informed weighting of vulnerability factors. This is especially advantageous in places like Chott Hodna that are only partially dry, where the environment changes a lot and there is not much data, making fixed-weight models less reliable. AHP lets you prioritize structured parameters based on expert opinion. Its fuzzy logic extension, Fuzzy AHP, makes it even more reliable by taking into account the uncertainty and ambiguity in these judgments. Studies like Bera et al. (2022) and Saravanan et al. (2023) have used AHP-enhanced DRASTIC in semi-arid areas of India and found that it does a better job of predicting vulnerability than the original model [18,20].

### 3.4. DRASTIC and DRASTIC-LU models

A frequently used method to assess the sensitivity of aquifers to pollution is the DRASTIC model. It is based on seven main factors influencing the risk of contamination. Using Eq. 1, the overall DRASTIC index (VI) was then calculated by integrating these weighted scores

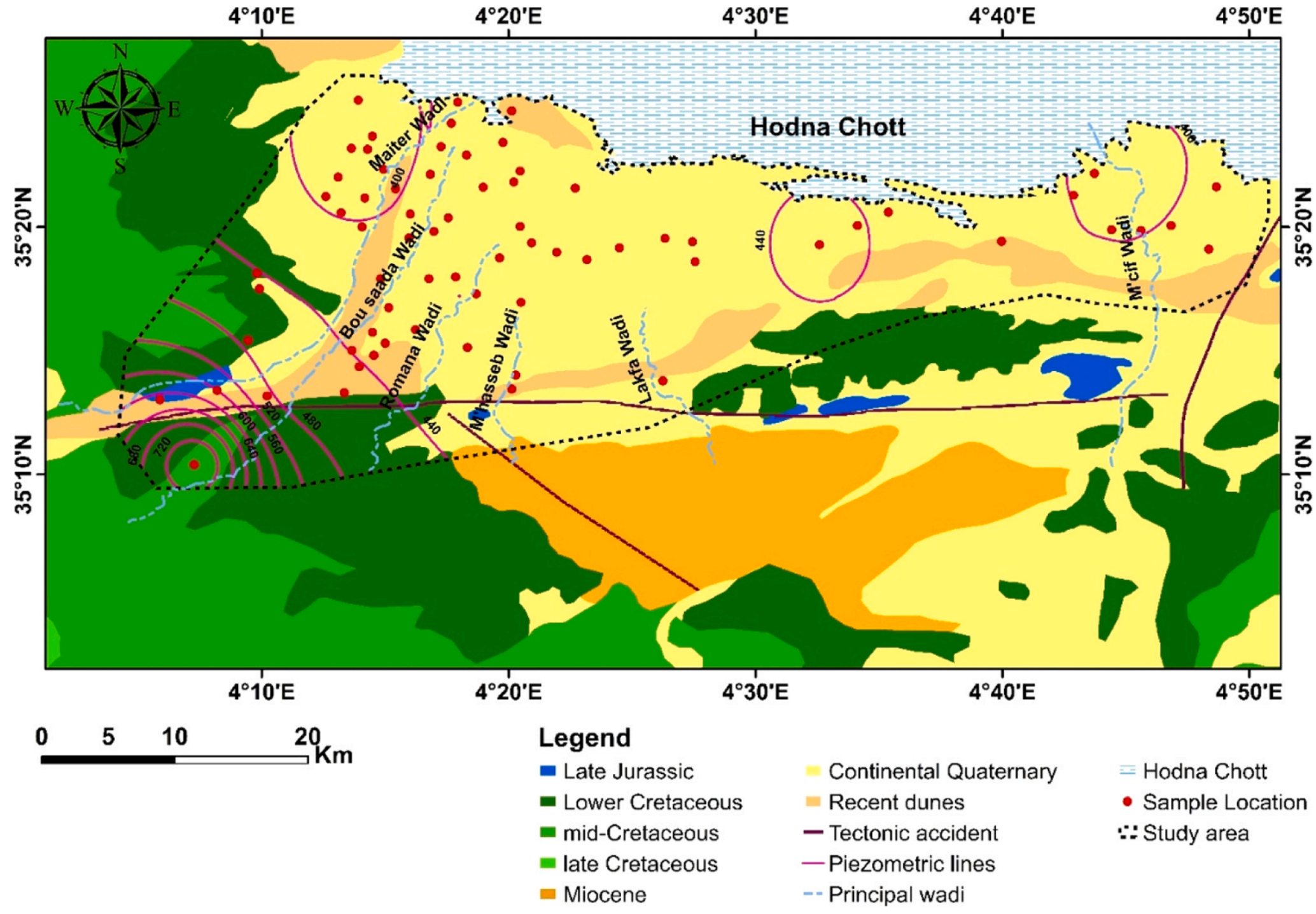

**Fig. 2.** Geological map of the study area with piezometric contours.

**Table 1**
Weight and rating coefficients of DRASTIC, modified DRASTIC_LU and AHP DRASTIC_LU [8,18–21].

| | *Range* | *Rating* | *DRASTIC Weight* | *DRASTIC_LU Weight* | *AHP DRASTIC_LU Weight* |
|---|---|---|---|---|---|
| **Depth to water (D) (m)** | 05.70–17.90 | 10 | 5 | 5 | 0.1786 |
| | 17.90–30.10 | 8 | | | |
| | 30.10–42.30 | 6 | | | |
| | 42.30–54.50 | 4 | | | |
| | 54.50–66.50 | 2 | | | |
| **Net recharge (R) (mm/year)** | 300–390 | 10 | 4 | 4 | 0.1429 |
| | 220–300 | 8 | | | |
| | 160–220 | 7 | | | |
| | 100–160 | 4 | | | |
| | 42–100 | 3 | | | |
| **Aquifer media (A) (dimensionless)** | Sand / Gravel | 8 | 3 | 3 | 0.1071 |
| | Sand and Clay | 6 | | | |
| | Limestone and Sandstone | 4 | | | |
| | Silty Clay | 2 | | | |
| **Soil media (S) (dimensionless)** | Pebble_Gravel_Sand | 10 | 2 | 2 | 0.0714 |
| | Gravel and Clay | 8 | | | |
| | Sand and Clay | 6 | | | |
| | Sandy Clay | 4 | | | |
| | Sandy Clay Loam | 3 | | | |
| **Topographic slope (T) (%)** | 0–2 | 10 | 1 | 1 | 0.0357 |
| | 2–6 | 8 | | | |
| | 6–11.50 | 6 | | | |
| | 11.50–19.50 | 4 | | | |
| | 19.50–40.50 | 2 | | | |
| **Impact of the vadose zone (I) (dimensionless)** | Pebble- Gravel-sand | 10 | 5 | 5 | 0.1786 |
| | Sandy Loam | 8 | | | |
| | Sandy Clay, Clayey Sand, Silt Loam, Loam | 6 | | | |
| | Clay Loam | 4 | | | |
| | Clay and Silty Clay | 2 | | | |
| **Hydraulic conductivity (C) (m/s)** | 4.15e−04–3.38 e−04 | 10 | 3 | 3 | 0.1071 |
| | 3.38e−04–2.61e−04 | 8 | | | |
| | 2.61e−04–1.84e−04 | 6 | | | |
| | 1.84e−04–1.06e−04 | 4 | | | |
| | 1.06e−04–2.96e−05 | 2 | | | |
| **Land use (LU) (dimensionless)** | Urban and Residential Areas | 10 | - | 5 | 0.1786 |
| | New agricultural Areas | 9 | | | |
| | Wastewater-Impacted Areas | 7 | | | |
| | Old agricultural Areas | 3 | | | |
| | Barren land | 1 | | | |

[22–25].

$$VI = D_rD_w + R_rR_w + A_rA_w + S_rS_w + T_rT_w + I_rI_w + C_rC_w \quad (1)$$

D: depth to the water table, R: net recharge rate of aquifer, A: aquifer media, S: soil types, T: topographic slope, I: impact of the vadose zone, C: hydraulic conductivity of aquifer, r: rating (score), and w: weight.

The DRASTIC model ignores anthropogenic factors and only considers geographical, geological, and climatological criteria. Integrating land use (LU) into the DRASTIC model improves the assessment by taking into account the impact of human activities on soils and groundwater, a particularly crucial aspect in areas with increasing human impact. Combining land use data with the DRASTIC model (called DRASTIC-LU) allows for concrete mapping of groundwater vulnerability. Eq. 2 was used to calculate the vulnerability index DRASTIC_LU.

$$VI_{LU} = D_rD_w + R_rR_w + A_rA_w + S_rS_w + T_rT_w + I_rI_w + C_rC_w + LU_rLU_w \quad (2)$$

Table 1 specifies the assignment of five weights to the LU factor. By incorporating the LU factor into Eq. (1), we calculate the vulnerability index of DRASTIC_LU, denoted by.

We calculated the DRASTIC Vulnerability Index (VI) by combining the weight and score assigned to each parameter. With 5 indicating the most important factor and 1 the least significant [26,27], the weights, which range from 1 to 5, reflect the relative importance of each parameter.

### *3.5. Analytic Hierarchy Process (AHP) approach*

The Analytic Hierarchy Process (AHP), introduced by Saaty [28]. is a multi-criteria decision-making (MCDM) tool designed to derive priority weights for multiple interrelated factors [29]. In groundwater vulnerability modeling, AHP is used to refine DRASTIC and DRASTIC_LU models by quantifying the relative influence of each parameter through pairwise comparisons.

Each parameter is compared against others using a 1–9 intensity scale (see Table 2) [28]. forming a reciprocal matrix $a_{ij}$:

**Table 2**
Analytical Hierarchy Process evaluation scale (Saaty 1977) [28].

| Intensity of importance | Definition | Explanation |
|---|---|---|
| 1 | Equal importance | Both parameters are of equal importance |
| 3 | Moderate importance | Parameter 1 is more important than parameter 2 |
| 5 | Strong importance | Parameter 1 is much more important than parameter 2 |
| 7 | Very strong importance | Parameter 1 has a very strong importance than parameter 2. |
| 9 | Extreme importance | Parameter 1 has absolute superior importance than parameter 2 |
| 2, 4, 6, 8 | Intermediate values | 2, 4, 6, 8 intermediate values |

$$[a_{ij}]_{N\times N} = \begin{bmatrix} 1 & a_{12} & \cdots\cdots\cdots a_{1N} \\ \frac{1}{a_{12}} & 1 \cdots\cdots\cdots a_{2N} \\ . & . & \cdots\cdots\cdots & . \\ . & . & \cdots\cdots\cdots & . \\ . & . & \cdots\cdots\cdots & . \\ \frac{1}{a_{1N}} & \frac{1}{a_{2N}} \cdots\cdots\cdots 1 \end{bmatrix} \quad i,j = 1, \ 2, \ldots N \tag{3}$$

where $N$ represents the number of parameters, and $a_{ij}$: the significance level of factor $i$ relative to factor $j$.

Dividing each element by the sum of its column normalizes the pairwise comparison matrix, so that the sum of the weights is equal to 1 (Eq. 4), where the normalized weight of each factor is $\omega_{ij}$.

$$\omega_{ij} = \frac{a_{ij}}{\sum a_{ij}} \tag{4}$$

Assessing the consistency of the results obtained from the pairwise comparison matrices is essential during the final stage of AHP deployment. Calculated using the approach of Saaty [28], this consistency index (CI) is (Eq. 5).

$$CI = \frac{(\lambda_{\max} - N)}{(N-1)}, CR = \frac{CI}{RI} \tag{5}$$

Where $\lambda_{\max}$ is the principal eigenvalue of the comparison matrix. After determining CI, the consistency ratio (CR) is then computed to evaluate the reliability of the judgments. $RI$ is the Random Index (1.41 for 8 parameters). Acceptable consistency is indicated by a CR $\leq$ 0.1 [30, 31].

This structured weighting allows for the generation of a refined vulnerability index within the AHP-optimized DRASTIC_LU model, thereby improving the mapping of important groundwater areas in semi-arid regions where data are scarce. To determine groundwater vulnerability in a semi-arid area, for example, weights may prioritize water table depth and recharge rates, as these factors are more important under such conditions.

### 3.6. FUZZY AHP approach

Particularly when data are incomplete or imprecise, a common challenge in semi-arid hydrogeological assessments [32,33], the fuzzy analytic hierarchy process (Fuzzy AHP) improves on traditional AHP by including fuzzy logic to address uncertainty in expert judgments. Instead of using crisp values, Fuzzy AHP uses Triangular Fuzzy Numbers (TFNs) defined as:

$$\widetilde{a} = (l, m, u)$$

$$\mu_i(x, \ l, \ m, \ u) = \begin{cases} 0 & x \leq l \\ \frac{x-l}{m-l} & if \ l < x \leq m \\ \frac{u-x}{u-m} & if \ m < x < u \\ 0 & if \ x \geq u \end{cases} \tag{6}$$

Where $l$ is lower bound (minimum judgment), $m$ is middle value (most likely judgment) and $u$ is upper-bound (maximum judgment).

In Fuzzy AHP, expert comparisons of criteria are expressed as fuzzy numbers rather than precise values [34]. The fuzzy pairwise comparison matrix is defined as:

$$\widetilde{A} = [\widetilde{a}_{ij}] = [(l_{ij}, m_{ij}, u_{ij})] \tag{7}$$

$\widetilde{a}_{ij}$ is the fuzzy comparison of criterion $i$ to criterion $j$. The fuzzy synthetic extent value $S_i$ for each criterion:

$$S_i = \frac{\sum_{j=1}^{n} \widetilde{a}_{ij}}{\sum_{i=1}^{n} \sum_{j=1}^{n} \widetilde{a}_{ij}} \tag{8}$$

Where $n$ is the number of criteria. This value represents the overall weight of each criterion.

After obtaining the fuzzy synthetic values, defuzzification is necessary to convert fuzzy numbers into crisp values. One common method is the Centroid Method or Center of Gravity (COG), which calculates the crisp value $C(\widetilde{a})$ of a fuzzy number $\widetilde{a} = (l, m, u)$ as:

$$C(\widetilde{a}) = \frac{l+m+u}{3} \tag{9}$$

This process allows for a stronger and more accurate way to determine weight, which better shows the uncertainty in hydro-environmental systems [35,36]. When integrated into the DRASTIC_LU framework, Fuzzy AHP produces finer-grained vulnerability maps, offering greater decision-making support for sustainable groundwater management in sensitive zones.

## 4. Results and discussion

### 4.1. Spatial characterization of DRASTIC parameters

Fig. 3 illustrates the locations of the eight DRASTIC parameters used to measure groundwater vulnerability in the southern Chott Hodna region. The depth to the water table (a) is lower in the west, which means it is more vulnerable. Net recharge (b) changes a lot, and there are areas along the wadi systems where a lot of water flows in. The aquifer (c) and soil (d) media maps show that the central and northern areas have permeable lithologies like sand, gravel, and sandy loam that make it easier for contaminants to move around. The slope of the land (e) is usually gentle, which makes it easier for water to soak into the ground. The vadose zone impact (f) is made up of a mix of low-retention clays and high-transmission sandy layers, which affect how quickly contaminants break down. The western and southern parts have high hydraulic conductivity (g), which means that the aquifer is more permeable. Land use (h) shows that there is a lot of farming going on and that wastewater is affecting many areas, especially in the central and southern plains. This adds to anthropogenic stress. These maps show how varied the hydrogeological vulnerability is in different parts of the study area.

### 4.2. Groundwater vulnerability map

#### 4.2.1. DRASTIC methods

The standard DRASTIC method calculates the vulnerability index (VI) using Eq. 1, adhering strictly to the weight and rating coefficients outlined in Table 1. The assessment of the groundwater vulnerability map created by the two different methods, DRASTIC and DRASTIC_LU, shows both strong similarities and important differences in the results. The land cover layer appears as a critical parameter influencing groundwater vulnerability, highlighting the need to integrate this factor with the intention of adding, in particular, the regions exposed to entropic pollution. The analysis identifies land cover (LU), groundwater depth (D), and aquifer characteristics (I) as the most influential parameters affecting groundwater pollution potential.

The VI index generated by this method ranged from 84 to 156. Based on these values, the groundwater vulnerability map was classified into three classes: low, moderate, and high vulnerability (Fig. 4a). In contrast, the modified DRASTIC_LU method yielded index ($VI_{LU}$) values ranging from 89 to 201.

This method identified 5.93 % of the study area as highly vulnerable, mainly in areas with intensive agricultural activities. An analysis of the land use map revealed that the highly vulnerable class mainly encompassed orchards and farmland. Furthermore, 32.37 % of the area was classified as highly vulnerable, while the moderate vulnerability class accounted for the largest portion, at 59.54 %. Conversely, the low vulnerability class accounted for the smallest area, at only 2.17 % (Fig. 4b).

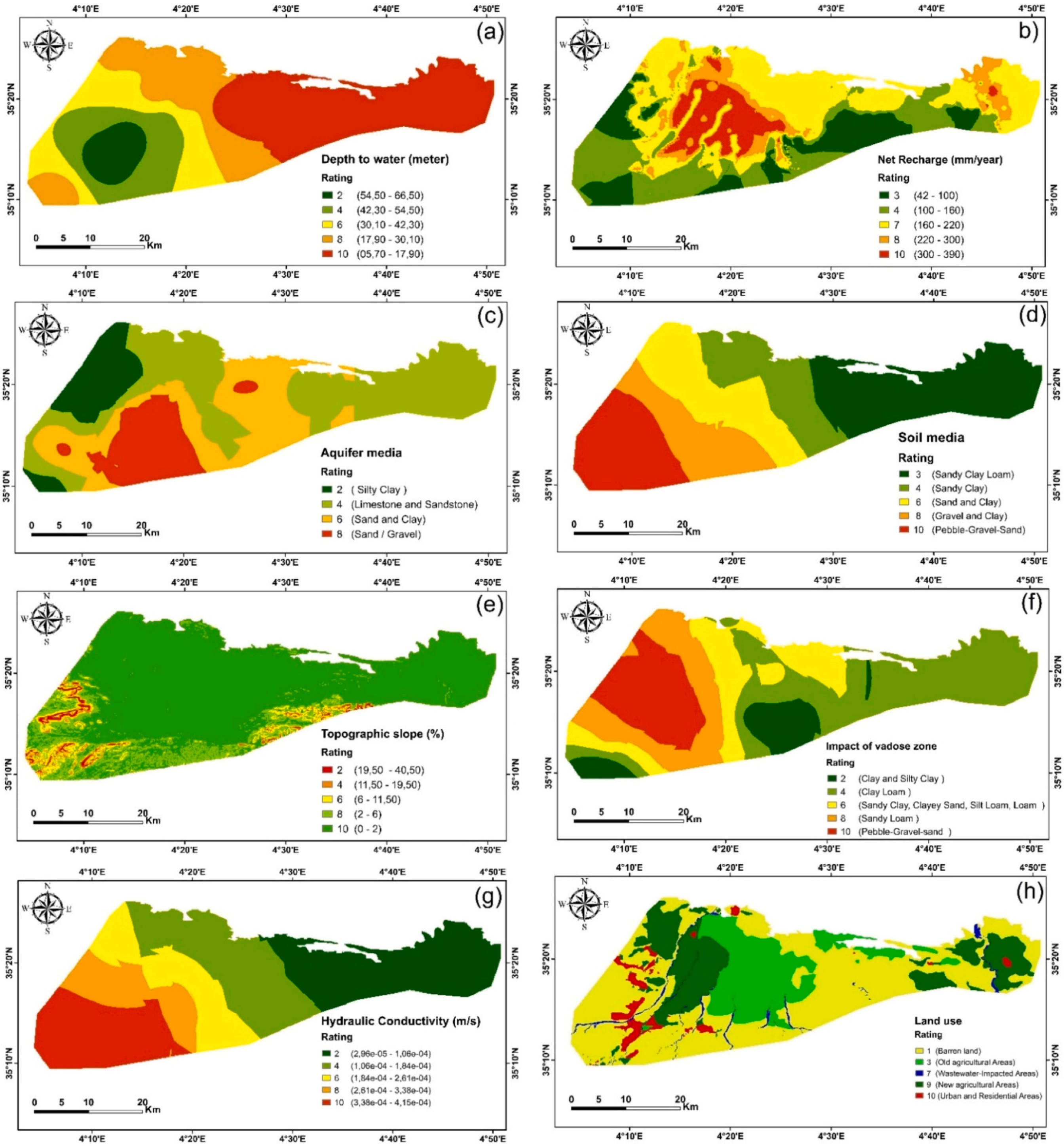

**Fig. 3.** Maps of groundwater vulnerability conditioning parameters:(a) depth to groundwater, (b) net recharge, (c) aquifer media, (d) soil media, (e) topography, (f) impact of the vadose zone, (g) conductivity of the aquifer, (h) land use.

According to the study, vulnerability is strongly influenced by groundwater depth, which ranges from 5.70 to 17.90 m; shallower groundwater levels are more sensitive to pollution. Higher net recharge rates (220–390 mm/year) also increase the risk of contaminants seeping into groundwater, thereby reducing pollution protection mechanisms. Generally, higher permeability, which increases vulnerability, corresponds to larger particle sizes in aquifers. This trend was particularly clear for sand and gravel, which had the highest vulnerability rates. Therefore, soil properties, such as its thickness, particle size, and expansion potential, strongly influence sensitivity to contamination. Soils with thicker layers and finer particles reduce vulnerability and enhance their ability to absorb contaminants. Topography also affects groundwater quality, thus affecting surface runoff dynamics. The results show that the highest leaching potential corresponds to a slope of 0–2 %, thus allowing the migration of pollutants into groundwater. The higher clay and gravel content also contributes to the reduction of groundwater permeability in the vadose zone, particularly in fine-grained soils. This work highlights the much greater impact of the vadose zone on sand and gravel than on other media.

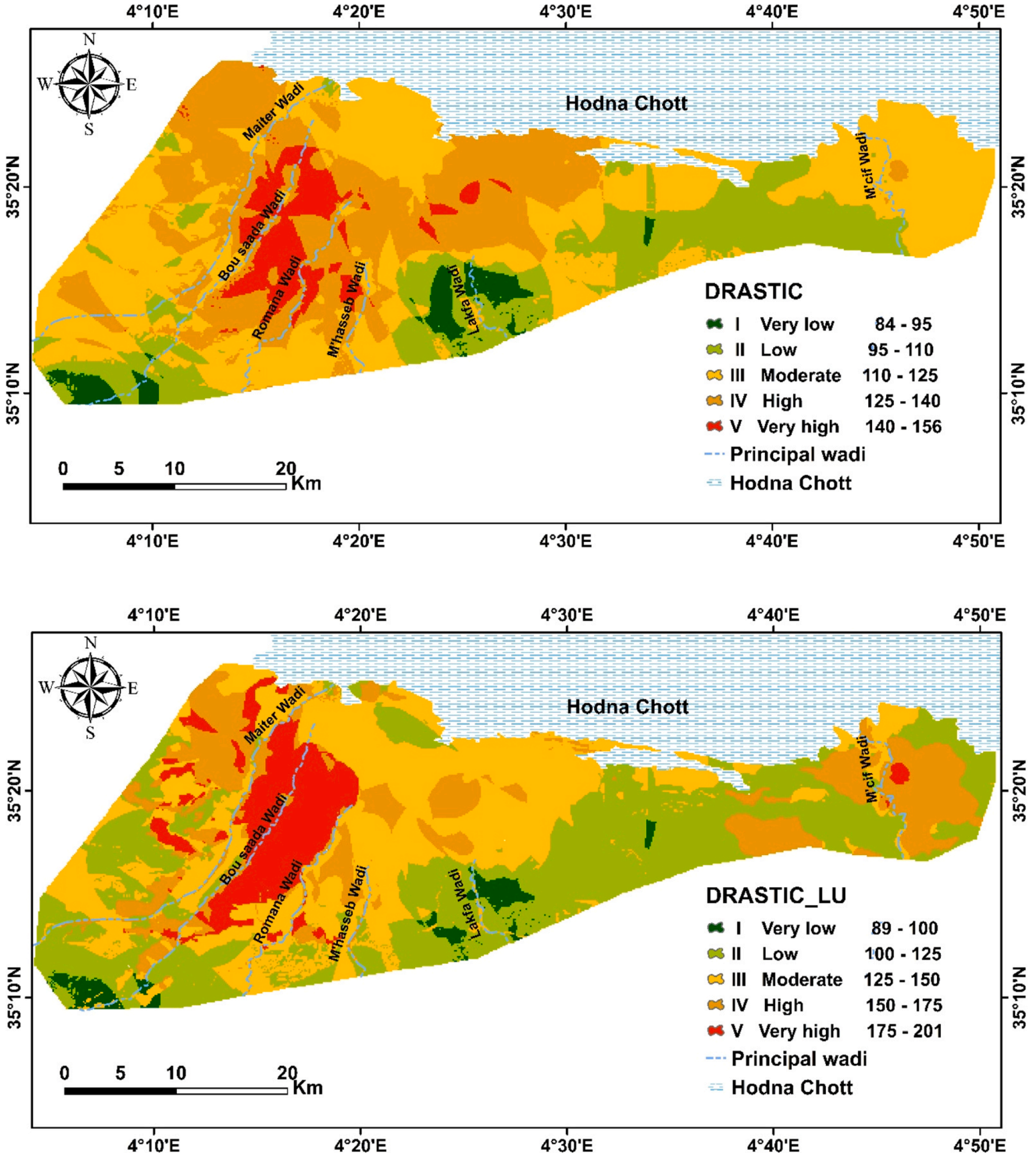

**Fig. 4.** Groundwater vulnerability maps: (a) DRASTIC, (b) DRASTIC_LU.

The application of the FUZZY_AHP method and its combination with DRASTIC_LU allows for a more precise assessment of contamination risk and groundwater vulnerability. Professionals can modify parameter weighting in the AHP, a multi-criteria approach [37], based on the relative relevance they assign. By allowing each element to be modified according to its relevance in the study area, the AHP contributes to a more precise assessment; for example, by giving greater importance to the depth of the water table in dry regions and to the relief in areas rich in hills and valleys.

#### 4.2.2. *AHP DRASTIC_LU method*

Using a pairwise comparison matrix from the Analytic Hierarchy Process (AHP), the AHP DRASTIC_LU method adds another parameter with established weighted priorities (see Tables 3 and 4).

The total weight calculated in the AHP DRASTIC_LU approach is the product of the AHP weights and the corresponding rating values.

Calculated alongside the seven initial parameters, the revised AHP

**Table 3**
Binary comparison matrix and weighted and weighted priorities of DRASTIC_LU method parameters.

| | D | R | A | S | T | I | C | LU | Weight AHP | Weight FUZZY-AHP |
|---|---|---|---|---|---|---|---|---|---|---|
| D | 1 | 5/4 | 5/3 | 5/2 | 5 | 1 | 5/3 | 1 | 0.1786 | 0.1233 |
| R | 4/5 | 1 | 4/3 | 2 | 4 | 4/5 | 4/3 | 4/5 | 0.1429 | 0.1169 |
| A | 3/5 | 3/4 | 1 | 3/2 | 3 | 3/5 | 1 | 3/5 | 0.1071 | 0.1142 |
| S | 2/5 | 1/2 | 2/3 | 1 | 2 | 2/5 | 2/3 | 2/5 | 0.0714 | 0.1207 |
| T | 1/5 | 1/4 | 1/3 | 1/2 | 1 | 1/5 | 1/3 | 1/5 | 0.0357 | 0.1640 |
| I | 1 | 5/4 | 5/3 | 5/2 | 5 | 1 | 5/3 | 1 | 0.1786 | 0.1233 |
| C | 3/5 | 3/4 | 1 | 3/2 | 3 | 3/5 | 1 | 3/5 | 0.1071 | 0.1142 |
| LU | 1 | 5/4 | 5/3 | 5/2 | 5 | 1 | 5/3 | 1 | 0.1786 | 0.1233 |

**Table 4**
Pairwise comparison matrix of DRASTIC_LU parameters using the AHP method.

| | D | R | A | S | T | I | C | LU |
|---|---|---|---|---|---|---|---|---|
| D | 1.00, 1.00, 1.00 | 1.25, 0.80, 1.00 | 1.67, 0.60, 1.00 | 2.50, 0.40, 1.00 | 5.00, 0.20, 1.00 | 1.00, 1.00, 1.00 | 1.67, 0.60, 1.00 | 1.00, 1.00, 1.00 |
| R | 0.80, 1.25, 1.00 | 1.00, 1.00, 1.00 | 1.33, 0.75, 1.00 | 2.00, 0.50, 1.00 | 4.00, 0.25, 1.00 | 0.80, 1.25, 1.00 | 1.33, 0.75, 1.00 | 0.80, 1.25, 1.00 |
| A | 0.60, 1.67, 1.00 | 0.75, 1.33, 1.00 | 1.00, 1.00, 1.00 | 1.50, 0.67, 1.00 | 3.00, 0.33, 1.00 | 0.60, 1.67, 1.00 | 1.00, 1.00, 1.00 | 0.60, 1.67, 1.00 |
| S | 0.40, 2.50, 1.00 | 0.50, 2.00, 1.00 | 0.67, 1.50, 1.00 | 1.00, 1.00, 1.00 | 2.00, 0.50, 1.00 | 0.40, 2.50, 1.00 | 0.67, 1.50, 1.00 | 0.40, 2.50, 1.00 |
| T | 0.20, 5.00, 1.00 | 0.25, 4.00, 1.00 | 0.33, 3.00, 1.00 | 0.50, 2.00, 1.00 | 1.00, 1.00, 1.00 | 0.20, 5.00, 1.00 | 0.33, 3.00, 1.00 | 0.20, 5.00, 1.00 |
| I | 1.00, 1.00, 1.00 | 1.25, 0.80, 1.00 | 1.67, 0.60, 1.00 | 2.50, 0.40, 1.00 | 5.00, 0.20, 1.00 | 1.00, 1.00, 1.00 | 1.67, 0.60, 1.00 | 1.00, 1.00, 1.00 |
| C | 0.60, 1.67, 1.00 | 0.75, 1.33, 1.00 | 1.00, 1.00, 1.00 | 1.50, 0.67, 1.00 | 3.00, 0.33, 1.00 | 0.60, 1.67, 1.00 | 1.00, 1.00, 1.00 | 0.60, 1.67, 1.00 |
| LU | 1.00, 1.00, 1.00 | 1.25, 0.80, 1.00 | 1.67, 0.60, 1.00 | 2.50, 0.40, 1.00 | 5.00, 0.20, 1.00 | 1.00, 1.00, 1.00 | 1.67, 0.60, 1.00 | 1.00, 1.00, 1.00 |

DRASTIC_LU index improved the groundwater vulnerability map. This approach divides the index values into four vulnerability categories. As with the DRASTIC_LU technique, agricultural areas are the most highly vulnerable. Of the total area, the "high vulnerability" category represents 32.49 %, while the "moderate vulnerability" category remains predominant at 58.53 %. The "low vulnerability" class makes up 3.18 % of the total area.

#### 4.2.3. FUZZY_AHP DRASTIC_LU method

Using the FUZZY AHP method with DRASTIC_LU makes groundwater vulnerability assessments more accurate, especially in dry areas like the Chott Hodna region (Fig. 5b), where things can change without warning and some data may not be completely correct. This method cuts down on mistakes that come from strict weighting and imprecise data. The FUZZY AHP method makes it easier to make decisions, makes it easier to move between vulnerability zones, and handles uncertainty in expert judgments. This model dealt with both uncertainties about expert data and quantitative variability better than previous ones.

Table 5 presents the percentage distribution of groundwater vulnerability classes derived from four modeling approaches. Compared with the original DRASTIC model, the integration of land use (DRASTIC_LU) and the application of multi-criteria decision support techniques (AHP and FUZZY_AHP) resulted in notable changes in vulnerability classification. The AHP-enhanced model, on the other hand, has a higher percentage of areas that are classified as "low" and "very low" vulnerability, which suggests that it is better at telling apart areas that are less affected. On the other hand, the FUZZY_AHP variant brought back a wider "moderate" class, which shows that it can better manage the uncertainty and subtle differences in semi-arid environments.

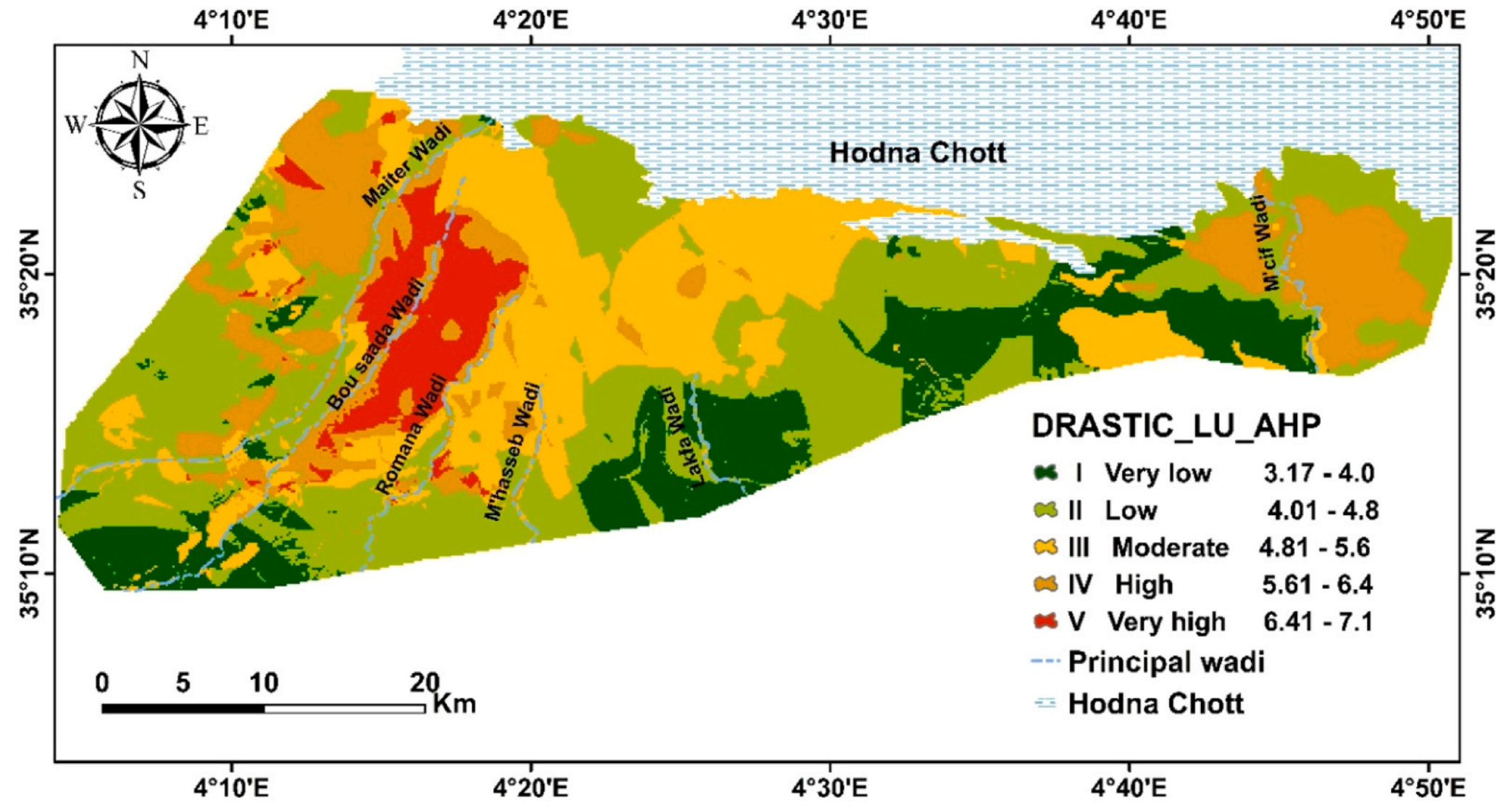

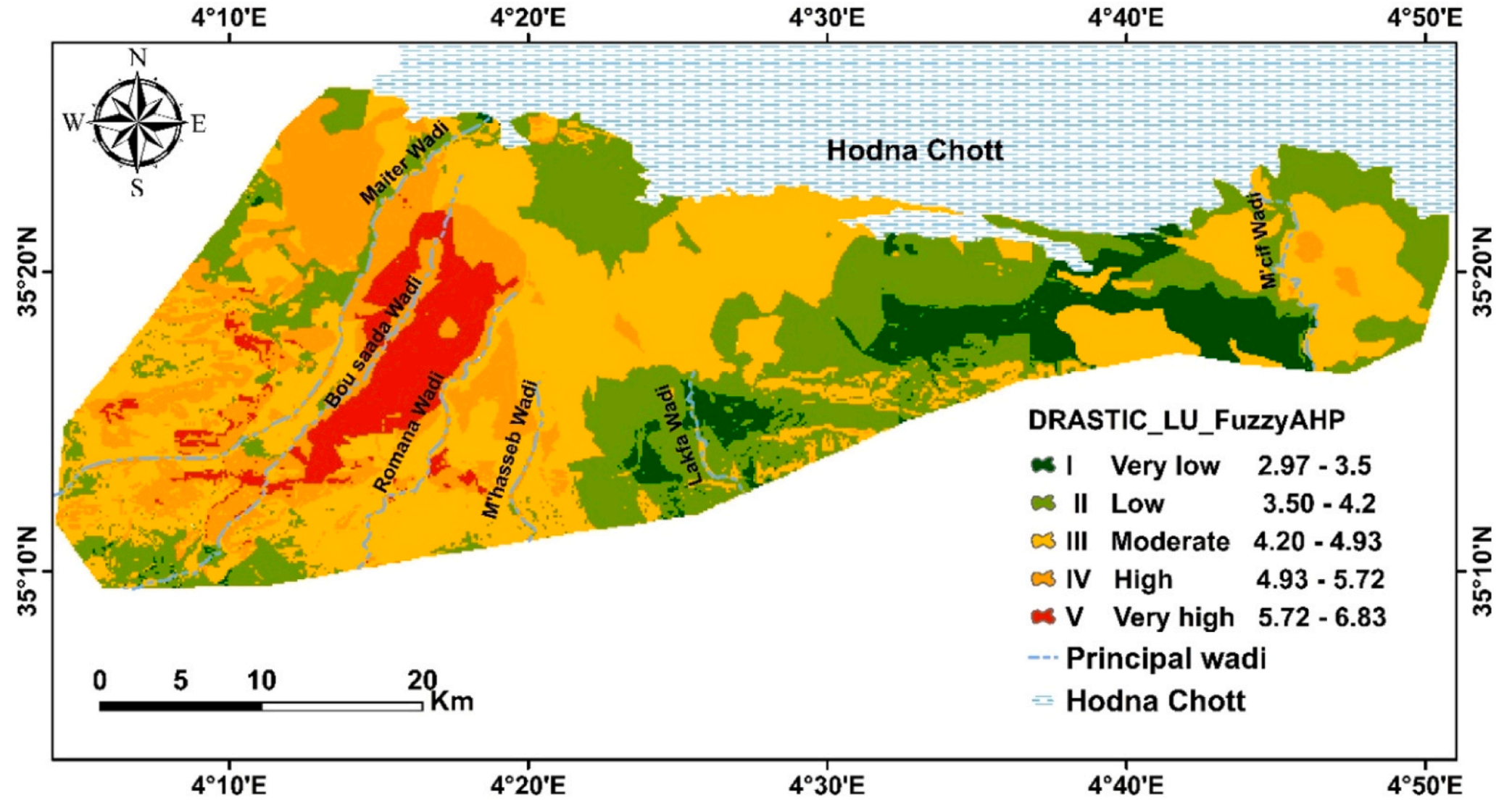

**Fig. 5.** Groundwater vulnerability maps: (a) AHP DRASTIC_LU, (b) FUZZY_AHP DRASTIC_LU.

**Table 5**
Comparative Areal (%) of Vulnerability Classes in different methods.

| Vulnerability class | Area (%) | | | |
|---|---|---|---|---|
| | DRASTIC | DRASTIC-LU | AHP DRASTIC_LU | FUZZY_AHP DRASTIC_LU |
| Very low | 3.20 | 2.17 | 15.39 | 7.41 |
| Low | 19.70 | 35.39 | 37.68 | 28.50 |
| Moderate | 42.22 | 33.93 | 23.93 | 41.69 |
| High | 28.88 | 18.75 | 17.10 | 16.27 |
| Very high | 5.99 | 9.75 | 5.90 | 6.12 |

### *4.3. Validation of methods*

Validation of the proposed groundwater vulnerability models was carried out by comparing the vulnerability indices (VI) generated by methods against nitrate concentrations measured in 70 wells (see Fig. 6a). The spatial validation confirms that integrating land use data and applying expert-optimized weights significantly enhances model performance. In the original DRASTIC model, 62.07 % of nitrate samples exceeding the WHO limit of 50 mg/L were located in high- to very-high-vulnerability zones. This detection rate improved to 82.76 % in the DRASTIC_LU model, 87.10 % in the AHP DRASTIC_LU model, and reached 88.41 % in the FUZZY_AHP DRASTIC_LU framework. Conversely, the proportion of samples with nitrate concentrations below 50 mg/L falling in low- to moderate-vulnerability zones declined from 80.49 % (DRASTIC) to 60.98 % (DRASTIC_LU), then slightly improved to 63.41 % (AHP DRASTIC_LU) and 64.93 % in the FUZZY-enhanced model.

We used Receiver Operating Characteristic (ROC) curve analysis [38] to check and compare how well the four models for assessing groundwater vulnerability predicted what would happen. This method of statistics checks how well a model works by plotting the true positive rate (sensitivity) against the false positive rate (1 – specificity) at different threshold levels. The Area Under the Curve (AUC) measures how well a model works: values close to 1 mean that the model is very effective at telling the difference between things, while values close to 0.5 mean that the model is doing random things. In this study, we made ROC curves by looking at the relationship between vulnerability index scores and nitrate levels in groundwater.

Fig. 6b shows that the original DRASTIC model had the lowest AUC score of 0.812, which means it wasn't very effective at predicting areas with nitrate contamination. Adding land use to the DRASTIC_LU model raised the AUC to 0.864. Adding AHP-based weighting refinement (AHP DRASTIC_LU) raised it even more to 0.875. The FUZZY_AHP DRASTIC_LU model had the highest predictive accuracy, with an AUC of 0.951.

Table 6 shows that as the model became more complex, the percentage of areas classified as having a high or very high risk of contamination decreased, while the alignment with actual nitrate contamination improved. This is in line with what has been written about recently. For example, Williams et al. (2024) found that adding land use and AHP to vulnerability models made them much more sensitive to nitrate pollution in dry areas [39]. Idir et al. (2024) also used ROC analysis to verify the accuracy of DRASTIC and other models in Algerian aquifers and stressed the need for hybrid frameworks [40]. Overall, these results back up our method and show how useful it is to combine DRASTIC with MCDA and fuzzy tools to better assess spatial risk and protect groundwater in areas that are at risk.

## 5. Discussion

Using two improved models, (a) AHP DRASTIC_LU and (b) FUZZY_AHP DRASTIC_LU, Fig. 7 shows spatial overlays of groundwater vulnerability and nitrate contamination across the Chott Hodna South region. Based on both natural and man-made factors, vulnerability zones are divided into five groups: very low (green), low (yellow), medium (orange), high (red), and very high (red). The pink dots show concentrations greater than 50 mg/L, while the blue dots show levels less than or equal to 50 mg/L. These points are very closely related to areas with high and very high vulnerability, especially in the western and north-western zones (around Bousaada and M'hasseb wadis).

These maps show that the improved models are better at predicting what will happen, especially the FUZZY_AHP DRASTIC_LU, which shows nitrate-affected areas more clearly. This alignment shows how important it is to use fuzzy logic, expert judgment, and land use together to get a favorable idea of how vulnerable groundwater is in semi-arid basins.

The classic DRASTIC model is a key tool for figuring out how vulnerable groundwater is on its own. But it can't fully take into account human-made pressures, which makes it less useful, especially in semi-arid areas like Chott Hodna South, where changes in land use have a big effect on how groundwater moves. Our findings confirm that integrating land use into the DRASTIC framework—via the DRASTIC_LU model—enhances both interpretability and practical relevance. The addition of Multi-Criteria Decision Analysis (MCDA) methods, especially Analytical Hierarchy Process (AHP) and its fuzzy logic variant (Fuzzy-AHP), further improves adaptability and decision-making reliability [33].

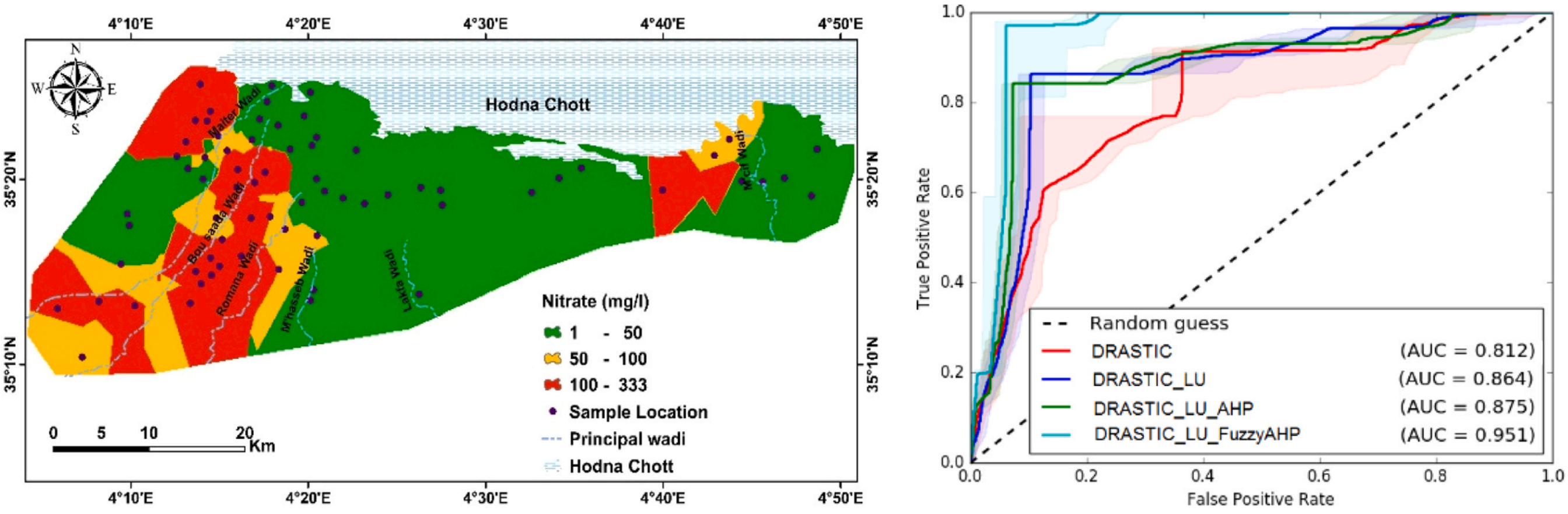

**Fig. 6.** a) Spatial distribution maps of $NO_3^-$ using ordinary Kriging b) ROC curve.

**Table 6**
Comparative Performance Metrics of DRASTIC, AHP, and Fuzzy AHP Approaches.

| Method | High + Very High Vulnerability Area (%) | Nitrate > 50 mg/L within High/Very High Zones (%) | Nitrate ≤ 50 mg/L within Low/Moderate Zones (%) | ROC AUC Score |
|---|---|---|---|---|
| DRASTIC | 34.87 % | 62.07 % | 80.49 % | 0.812 |
| DRASTIC_LU | 28.50 % | 82.76 % | 60.98 % | 0.864 |
| AHP DRASTIC_LU | 23.00 % | 87.10 % | 63.41 % | 0.875 |
| FUZZY_AHP DRASTIC_LU | 22.39 % | 88.41 % | 64.93 % | 0.951 |

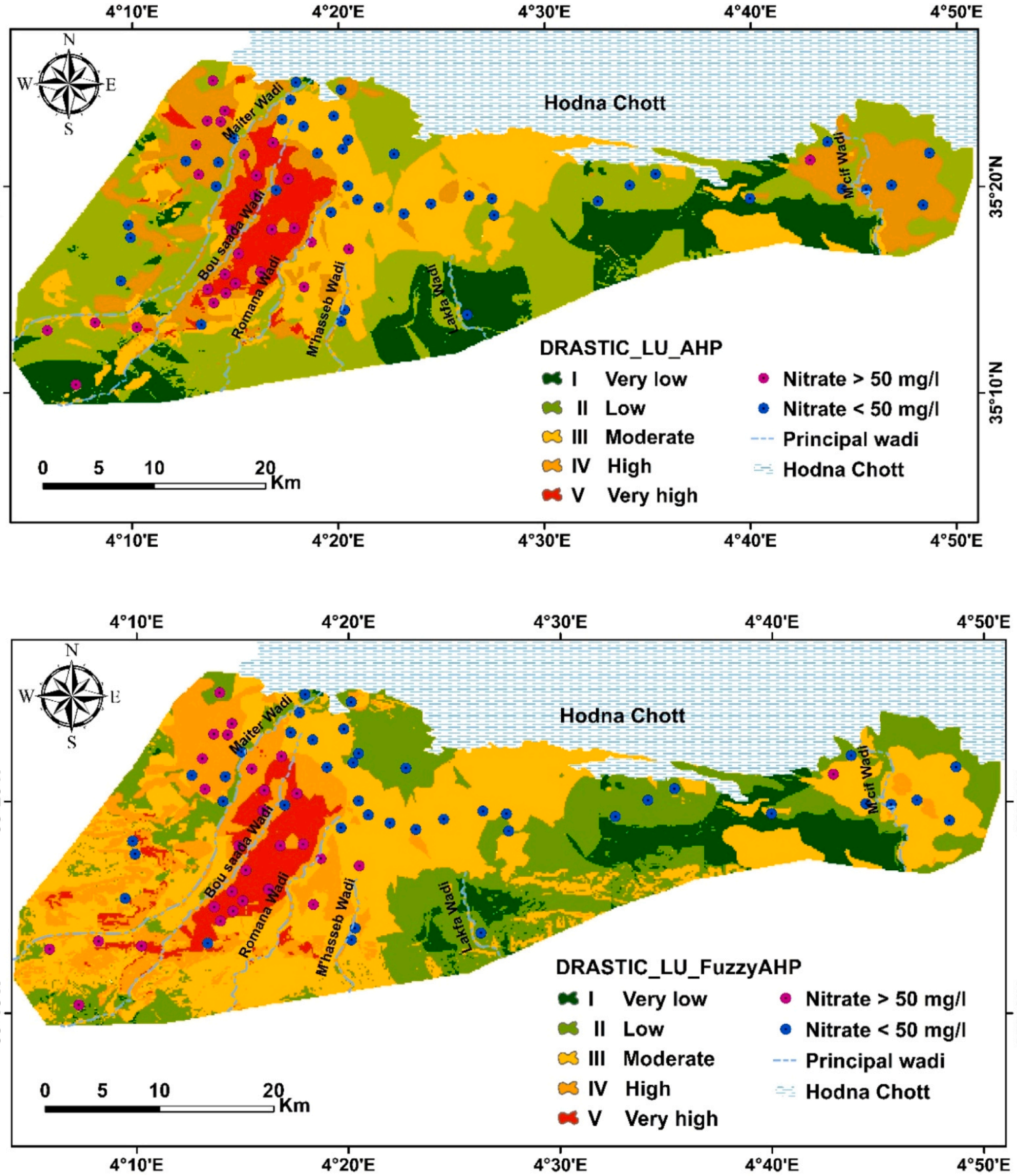

**Fig. 7.** Groundwater vulnerability maps: (a) AHP DRASTIC_LU, (b) FUZZY_AHP DRASTIC_LU compared to nitrate concentration.

The ROC-AUC scores—0.941 for DRASTIC_LU and 0.951 for FUZZY AHP DRASTIC_LU—demonstrate strong predictive alignment with measured nitrate levels, a performance echoed by recent works. For instance, Williams et al. (2024) showed that integrating land use and climate factors into modified DRASTIC models substantially improved vulnerability classification accuracy in the semi-arid Verde River Basin [39]. In the same way, Yuan et al. (2024) used dual-method validation and high AUC values to confirm changes in Zhengzhou, China's vulnerability that were caused by land use [41]. Całka et al. (2025) used Fuzzy-AHP to make land-use planning decisions less subjective and showed that spatial assessments were more reliable, which supports the idea that fuzzy logic is effective at modeling uncertainty [42]. These methods are in line with our findings, especially in semi-arid areas where there aren't enough data and the terrain is very different.

Our comparison shows that as the methods get more complicated, the area that is considered very vulnerable gets smaller, but the match with real nitrate data gets better. This trend shows that there is better spatial resolution and less overgeneralization. In our comparison of weights, AHP gave the highest normalized weights (0.1786) to land use, depth to water, and vadose zone impact. These are all indicators of both human influence and hydrogeological sensitivity. FUZZY AHP put even more emphasis on topographic slope (0.1640) and soil media (0.1207). It did this by showing how different variables are related to each other in a flexible, data-driven way. Despite these improvements, reliance on expert judgment in AHP-based models remains a source of subjectivity. While consistency ratio tests were applied, future research should explore participatory or automated weighting to enhance reproducibility. The data also limits the model's accuracy, particularly when it comes to land use changes and pollution levels. However, this study provides a robust and flexible method to assess the vulnerability of groundwater in dry and semi-dry regions. By integrating land use and decision support into vulnerability mapping, the approach makes progress toward global goals in climate-resilient land planning, safe drinking water access, and sustainable water management. Furthermore, by addressing groundwater stress in a semi-arid region, the research advances Climate Action, providing a framework for adapting water management practices in the face of increasing environmental pressures.

## 6. Conclusion

By improving the traditional DRASTIC method, this study provides a full picture of how vulnerable groundwater is in the semi-arid South Chott Hodna region. The models are more accurate and flexible for local conditions because they add land use as a factor (DRASTIC_LU) and fine-tune parameter weights using multi-criteria decision-making tools like the Analytical Hierarchy Process (AHP) and its fuzzy logic version (FUZZY AHP).

We checked the models against nitrate levels measured in the field. The ROC-AUC scores went up from 0.812 in the original DRASTIC model to 0.875 (AHP DRASTIC_LU) and 0.951 (FUZZY AHP DRASTIC_LU), which shows that the models got a lot better at predicting. Also, the percentage of high nitrate concentrations (>50 mg/L) in areas with high to very high vulnerability rose from 62.07 % (DRASTIC) to 88.41 % (FUZZY AHP DRASTIC_LU). This improvement demonstrates a closer alignment between the model outputs and contamination patterns. The comparison of parameter weights shows that MCDA methods (AHP and FUZZY AHP) offer more flexible and context-sensitive evaluations than the fixed integer weights used in traditional models. Land use, depth to water, and vadose zone impact were the most important factors, especially in the AHP scheme. Slope and soil media became more important in the fuzzy framework. The study does, however, point out some problems with these improvements: it only used nitrate as a validation parameter, the hydrogeological data varied by location, and the weights given by experts were subjective. Future studies should look into multi-contaminant validation, include temporal variability, and use data-driven or participatory weighting methods to reduce bias.

In conclusion, using DRASTIC variants, GIS, and MCDA techniques together shows a lot of promise as a tool to help make decisions about how to manage groundwater sustainably in dry and semi-dry areas. It helps locate vulnerable areas more accurately, encourages smart land-use planning, and is in line with global goals for sustainability, especially when it comes to protecting water resources and making agriculture more resilient to climate change and more sustainable.

## CRediT authorship contribution statement

**Messaoud Ghodbane:** Software, Formal analysis. **Tahar Selmane:** Software. **Lakhdar Seraiche:** Writing – original draft, Software. **Dougha Mostefa:** Writing – review & editing, Supervision. **Ahmed Ferhati:** Visualization, Data curation. **Djamal Eddine Djemiat:** Visualization, Software.

## Funding

This work is supported by the Algerian Ministry of Higher Education and Scientific Research and the General Directorate of Scientific Research and Technological Development (DGRSDT). The funders had no role in study design, data collection and analysis, the decision to publish, or manuscript preparation.

## Declaration of Competing Interest

The authors declare no conflicts of interest.

## Declaration of Competing Interest

Authors declare that there are no known competing financial interests or personal relationships that could have appeared to influence the work reported in this manuscript titled.

## Acknowledgements

The authors express gratitude to the Ministry of Higher Education and Scientific Research of Algeria for providing partial financial support, enabling the successful completion of the research. Without this support, the undertaking would have been beyond our means.

## Data Availability

Data will be made available on request.